\documentclass[preprint2]{aastex}

\received{2001 December 17}
\begin{document}

\title{Non-Gaussian Features of Transmitted Flux of QSO's 
       Ly$\alpha$ Absorption: Intermittent Exponent}

\author{Jesus Pando\altaffilmark{1},        
       Long-Long Feng\altaffilmark{2,3}, 
       Priya Jamkhedkar\altaffilmark{4},
       Wei Zheng\altaffilmark{5},
       David Kirkman\altaffilmark{6}, 
       David Tytler\altaffilmark{6}
       and Li-Zhi Fang\altaffilmark{4}  }

\altaffiltext{1}{Department of Physics, DePaul University, Chicago Il, 60614 }
\altaffiltext{2}{Center for Astrophysics, University of Science
and Technology of China, Hefei, Anhui 230026,P.R.China}
\altaffiltext{3}{National Astronomical Observatories, Chinese
Academy of Science, Chao-Yang District, Beijing, 100012, P.R.
China} 
\altaffiltext{4}{Department of Physics, University of Arizona,
Tucson, AZ 85721}
\altaffiltext{5}{Department of Physics and Astronomy, Johns 
   Hopkins University, MD 21218-2686}
\altaffiltext{6}{Center of Astrophysics and Astronomy, University of
          California, San Diego, CA 92093}

\begin{abstract}

Recently, it has been found that the field traced by QSO's Ly$\alpha$ 
forests is intermittent on small scales. Intermittent behavior is essential 
for 
understanding the statistics and dynamics of cosmic gravitational 
clustering in the nonlinear regime. The most effective method of 
describing intermittency uses the structure functions and the intermittent 
exponent, which measure the scale- and order-dependencies of 
the ratio between the higher order to second order moments of the field. 
These properties can be used not only to confirm the non-gaussianity of  
fields, but also to detect the {\it type} of non-gaussianity.

In this paper, we calculate the structure function and intermittent 
exponent of 1.) Keck data, which consists of 28 high resolution, high 
signal to noise ratio (S/N) QSO Ly$\alpha$ absorption spectra, and 2.)
Ly$\alpha$ forest simulation samples produced via the pseudo hydro scheme 
for the low density cold dark matter (LCDM) model and warm dark matter (WDM) 
model with particle mass $m_W=300, 600, 800$ and 
1000 eV. Aside from the WDM model with $m_W=300$ eV, the simulation samples 
are in agreement with observations in the context of the power 
spectrum. We find, however, that the intermittent behavior of all the 
simulation samples 
is substantially inconsistent, both quantitatively and qualitatively, with 
the Keck data. Specifically, 1.) the structure functions of 
the simulation samples are significantly larger than that of Keck data 
on scales $k \geq 0.1$ km$^{-1}$ s, 2.) the intermittent 
exponent of the simulation samples is more negative than that of Keck 
data on all redshifts considered, 3.) the order-dependence of the
structure functions of simulation samples are closer to the intermittency
of hierarchical clustering on all scales, while the Keck data are  
closer to a lognormal field on small scales. These differences are independent
of noise and show that the intermittent evolution modeled by the 
pseudo-hydro simulation is substantially different from observations, even 
though they are in good agreement with each other in terms of second and 
lower order statistics. This 
result also shows that ``weakly" clustered samples, like high resolution 
Ly$\alpha$ absorption spectrum, are effective in testing dynamical models of 
structure formation if their intermittent features are considered.

\end{abstract}


\section{Introduction}

The non-gaussian features of the QSO's Ly$\alpha$ transmitted 
flux were recently studied and showed that the cosmic mass field traced 
by the QSO's Ly$\alpha$ forests is intermittent on small scales. That is, the 
probability distribution functions (PDF) of the transmitted flux 
fluctuations were found to be substantially long tailed on scales less 
than $\sim$ 1 h$^{-1}$ Mpc. The spatial distribution of the power of the 
transmission fluctuations was found to be spiky, i.e., the power was 
concentrated in rare modes while most modes showed very low power (Jamkhedkar, Zhan 
and Fang 2000; Zhan, Jamkhedkar and Fang 2001; Feng, Pando and Fang 2001).

The impact of intermittency on observational cosmology is 
that the power spectrum is no longer effective for testing models 
when intermittency is substantial. Although spikes in the intermittent 
field are rare and improbable events, one cannot neglect the spikes
in measuring the power spectrum as almost all power of the transmission 
fluctuations is concentrated in these events. In such a case, the power 
spectrum is dominated by rare and improbable events (spikes). An 
intermittent field  is statistically homogeneous. Yet, the rare events 
lead to significant differences among samples from different parts of 
the universe when the spatial size of the region is not large enough to 
contain numerous spikes. The precision of determining the power spectra 
is intrinsically constrained by intermittency. The uncertainty of the 
power spectrum for an intermittent field is inevitably large (Jamkhedkar 
et al 2002, hereafter Paper I). 

Therefore, in the nonlinear regime, more effective measures to
compare the predictions of models and observations must be used. 
Although an intermittent
field is non-gaussian, many non-gaussian statistics are ineffective or
insensitive to intermittency. This is because we need not only
techniques for confirming the non-gaussianity of a field, but also methods
which can tell the {\it type} of non-gaussianity. The most effective 
measures sensitive to the {\it type} of intermittency are 
the structure functions and the intermittent exponent (e.g. 
Shraiman \& Siggia 2000). 

In this paper, using a set of 28 high resolution, high signal to noise ratio 
(S/N) QSO Ly$\alpha$ absorption spectra (Keck data), we will carry out a 
systematic investigation of the structure functions and the 
intermittent exponent of the transmitted flux. To show the usefulness of the 
structure function and intermittent exponent as model discriminators, we also
study the intermittent behaviors of Ly$\alpha$ forest samples produced by
a pseudo-hydro simulation technique for the LCDM and WDM models. These models 
are among the best that match the observed power spectrum and other lower 
than second order statistics. We will show that the simulation samples are 
significantly intermittent. However, both the quantitative and qualitative 
properties of simulation samples are found to be substantially different 
from the Keck data. 
 
The paper is organized as follows. \S 2 addresses the method for measuring
intermittent random fields. \S 3 presents the algorithm for calculating 
the structure functions and intermittent exponent. \S 4 describes, 
both the observed and simulated samples, that we will use. The power spectrum 
and other lower order statistics of the Keck data and simulation 
samples are given in \S 5. \S 6 contains the important result, the 
structure function, intermittent exponent and their of order-, redshift-
and scale-dependencies for the Keck data and simulation samples. 
The emphasis is on revealing the deviation of the intermittent behavior for 
the simulation samples from the Keck data. Finally, the conclusions and 
discussions are in \S 7.

\section{Statistical description of an intermittent random field} 

\subsection{Basic properties of an intermittent field}

Physically, intermittency is used to characterize a special type of 
random field in which structures are essentially strong enhancements, 
peaks or spikes, randomly and widely scattered in space and/or time, 
with a low field value between the spikes. The spike-gap-spike 
feature is more pronounced on smaller scales.  That is, compared 
with gaussian fields, the probability distribution function (PDF) of 
the fluctuations possesses a high peak around zero and possibly a long-tail.
 The long-tail events correspond to the rare and improbable high peaks.  
This feature was originally found in the temperature and velocity 
distributions in turbulence (Batchelor \& Townsend 1949). 

Mathematically, an intermittent random density field $\rho(x)$ is 
defined by the ratio between the high- and low-order moments of the 
field 
\begin{equation}
\frac{\langle [\rho(x+r)- \rho(x)]^{2n}\rangle}
{[\langle [\rho(x+r)- \rho(x)]^2\rangle]^n}
\asymp\left \langle \frac{r}{L} \right \rangle^{\zeta(n)}, 
\end{equation}
where $L$ is the size of the sample, and the exponent $\zeta$ can be 
function of $n$ and $r$. If $\zeta$ is negative for small $r$, the ratio 
eq.(1) is divergent with $r\rightarrow  0$. This corresponds to an intermittent
field. An intermittent field is characterized by the 
$n$- and $r$-dependencies of exponent $\zeta$. 

These properties make characterizing intermittent fields by traditional 
statistical measures difficult. The power spectrum and two-point correlation 
function are unable to quantify intermittency. Two fields 
having the same power spectrum may have very different tails in their
PDF.  Very different from the linear regime, 
the power spectrum will on longer be a critical discriminator among 
models of structure formation for an intermittent field.

Equally ineffective are any individual higher order correlation 
functions and higher order moments of the density fluctuations 
$\delta(x)=[\rho(x)-\bar{ \rho}]/\bar{\rho}$. Three- and four-point 
correlation functions of the density perturbation $\delta(x)$ are very 
useful for distinguishing a gaussian from a non-gaussian field. However, 
they are insensitive to the 
difference between an intermittent field and a non-gaussian, but 
non-intermittent field. Moreover, eq.(1) means that the PDF
of the density fluctuations $[\rho(x+r)- \rho(x)]$ cannot be 
expanded into a series of moments of $[\rho(x+r)- \rho(x)]$,
as it does not converge.  In this case, a better measure is not
the divergent quantity itself, 
$\langle [\rho(x+r)- \rho(x)]^{2n}\rangle/
\langle [\rho(x+r)- \rho(x)]^2\rangle ^n$ when $r\rightarrow  0$, 
but the quantity describing the divergent behavior, specifically,
the exponent $\zeta$. 

\subsection{Definition of the intermittent exponent}

Let us define density difference $\Delta_r(x)\equiv \rho(x+r)-\rho(x)$.
The density difference is equal to the density contrast difference,
$\delta(x+r)-\delta(x)$, if the mean density $\bar{\rho}$ is normalized 
to 1. Intermittency
is now defined by the divergence of the ratio between the high and low 
order moments $|\Delta_r(x)|^n$. For a gaussian 
field, the statistical properties of a field $\rho(x)$ and its difference
$\rho(x+r)-\rho(x)$ are the same, while they are different for an
intermittent field. 

The ensemble average of the moment $|\Delta_r(x)|^{2n}$ is
\begin{equation}
S^{2n}_r = \langle|\Delta_r(x)|^{2n} \rangle
\end{equation}
where $n$ is a positive integer. If the field is homogeneous, 
$S^{2n}_r$ is independent of $x$ and depends only on $r$. $S^{2n}_r$
is called the structure function. When the ``fair sample hypothesis" 
is applicable (Peebles 1980), the structure function can be 
calculated as the spatial average 
\begin{equation}
S^{2n}_r =\frac{1}{L}\int |\Delta_r(x)|^{2n}dx,
\end{equation}
where $L$ is the spatial range of the sample. When $n=1$, we have
\begin{equation}
S^2_r =\langle |\Delta_r(x)|^{2} \rangle.
\end{equation}
$S^2_r$ is the mean of the square of the density fluctuations at 
wavenumber $k\simeq 2\pi/r$, and therefore, the $r$-dependence of
$S^2_r$ actually is a different version of the ordinary power 
spectrum of the field.
     
The intermittent exponent $\zeta$ is defined
by\footnote{In turbulence, $S^{2n}_r/[S^{2}_r]^n$ is used to define
the so-called anomalous scaling (Shraiman \& Siggia 2000.)}
\begin{equation}
\frac{S^{2n}_r}{[S^{2}_r]^n} \propto
   \left(\frac{r}{L} \right )^{\zeta}.
\end{equation}
The intermittency of the field is effectively measured by the 
structure function and the intermittent exponent. 

For an intermittent field, the ratio of $S^{2n}_r$ to $[S^{2}_r]^n$ 
is larger for smaller $r$, and therefore, the exponent $\zeta$ is 
negative. The condition $S^{2n}_r \gg [{S^{2}_r}]^n$ on small scales
$r\ll L$ indicates that the field contains  ``abnormal'' events 
of large density fluctuations $|\Delta_r(x)|$ on scale $r$. Thus, 
the more negative the exponent $\zeta$ on smaller scales, the stronger the 
``abnormal'' events on smaller scales. This gives rise to the spiky structure 
of an intermittent field. Generally speaking, the intermittent exponent 
$\zeta$ measures the smoothness of the field: for positive $\zeta$, 
the field is smoother on smaller scales. If $\zeta$ is negative, the 
field is rough on small scales, and can even be singular.

\subsection{Examples of the intermittent exponent}

\noindent{1. A Gaussian field.}

If the homogeneous and isotropic field is gaussian, the structure 
function eq.(2) is
\begin{eqnarray}
S^{2n}_r  & = & \int_{-\infty}^{\infty}
P_{g}[\Delta_r(x)]|\Delta_r(x)|^{2n} d\Delta_r(x) \\ \nonumber
 & = & (2n-1)!!
   \left[ S^2_r\right ]^n,
\end{eqnarray}
where $P_{g}$ is the gaussian PDF of $\Delta_r(x)$.
Thus, if the mass density field is gaussian, we have
\begin{equation}
\frac{S^{2n}_r}{[S^{2}_r]^n} = (2n-1)!!.
\end{equation}
This ratio is independent of scale $r$, and therefore, the 
intermittent exponent $\zeta=0$.  

\noindent{2. A self-similar field}

For a self-similar field, the average of $\Delta_r(x)$ on different scales 
satisfies 
\begin{equation}
\langle|\Delta_r(x)|^{2n} \rangle =
\lambda^{2nh}\langle|\Delta_{\lambda r}(x)|^{2n} \rangle
\end{equation}
where $\lambda$ gives the scale factor from $r$ to $\lambda r$, 
and $h$ is the self-similar index. In this case, we have 
\begin{equation}
S^{2n}_r=\left (\frac{r'}{r}\right )^{2nh} S^{2n}_{r'},
\end{equation}
and therefore 
\begin{equation}
\frac{S^{2n}_r}{[S^{2}_r]^n}=\frac{S^{2n}_{r'}}{[S^{2}_{r'}]^n}.
\end{equation}
Therefore, $S^{2n}_r/[S^{2}_r]^n$ is independent of $r$. We have then
$\zeta=0$ and self-similar fields are not intermittent.

\noindent{3. Hierarchical clustering.}

Hierarchical clustering is popular in modeling the nonlinear clustering 
of galaxies. This scheme assumes that the correlation functions of the 
mass density can be described by the linked-pair approximation, i.e., 
the $n$-th irreducible correlation function $\xi_n$ is given by the 
two-point correlation function $\xi_2$ as $\xi_n = Q_n \xi_2^{n-1}$, 
where the hierarchical coefficient $Q_n$ is constant (White 1979). 

It has been shown that  a hierarchical clustered field is (Feng, Pando 
\& Fang 2001)
\begin{equation}
 \frac{S^{2n}_r}{[S^{2}_r]^n} \propto \left(\frac{r}{L} 
  \right )^{-(d-\kappa)(n-1)},
\end{equation}
where $d$ is the spatial dimension, and the coefficient $\kappa$ is a 
constant depending on the power law index of the power spectrum. In the 
case where $\kappa <d$, the field is intermittent with exponent 
\begin{equation}
\zeta \simeq  -(d-\kappa)(n-1).
\end{equation}
This is the simplest type of intermittency -- a mono-fractal with fractal
dimension $\kappa$ in $d$-dimensional space. A phenomenological model 
with hierarchical relations was developed by Soneira \& Peebles (1977). 
The model is essentially the same as the so-called $\beta$-model of the
intermittency in turbulence where a dimension $\kappa$ fractal distribution 
in $d$-dimensional space has the exponent $\zeta$ is given by
eq.(12) (Frisch 1995.)

\noindent{4. A Lognormal field}

Since the time of Hubble, the lognormal distribution has been used to model 
the PDFs of the cosmic mass and velocity field, such as the one-point 
distribution of the number of galaxies, velocity difference, angular 
momentum, etc (e.g. Yang et al. 2001b).  The lognormal model of the 
baryonic matter distribution 
(Bi \& Davidsen, 1997; Feng \& Fang 2000) has also been found to be in 
good agreement with all observed properties of the Ly$\alpha$ forests. 

For a lognormal field, the PDF of $\Delta_r(x)$ is given by
\begin{equation}
P[\Delta_r(x)] =\frac{1}{2^{3/2}\pi^{1/2}|\Delta_r(x)|\sigma(r)}
 \exp\left \{ -\frac{1}{2}
\left (\frac {\ln|\Delta_r(x)|-\ln \overline{|\Delta_r(x)|}}
{\sigma(r)}\right)^2  \right \},
\end{equation}
where the variance $\sigma(r)$ of $\ln|\Delta_r(x)|$ can be a function 
of the scale $r$. With eq.(13), we have (Vanmarcke 1983)
\begin{equation}
\frac{S^{2n}_r}{[S^{2}_r]^n} = e^{2(n^2-n)\sigma^2(r)}.
\end{equation}
Using eq.(5), the intermittent exponent of a lognormal field is
\begin{equation}
\zeta \simeq 2(n^2-n)\sigma^2(r)/\ln(r/L).
\end{equation}
When $r < L$, $\zeta$ is negative. Therefore, a lognormal field is 
intermittent.

To summarize, the structure function and intermittent exponent provide 
a complete and unified description of intermittent fields. The 
$n$- and $r$-dependencies of the structure functions and intermittent 
exponent $\zeta$ are sensitive to the details of the intermittency
of the field. These measures are very powerful for distinguishing among 
fields that are gaussian, self-similar, mono-fractal, multi-fractal, and 
singular.

\section{The intermittent exponent}

\subsection{Statistical variables in the DWT representation}

The quantity $\Delta_r(x)$ or $[\delta(x+r)-\delta(x)]$ contains 
two variables: the position $x$ and the scale $r$, and therefore, 
$\delta(x)-\delta(x+r)$ should be calculated with a space-scale 
decomposition. Moreover, as defined by eq.(2), $S_r^{2n}$ 
is sometimes a diverging quantity. For instance, 
$\langle S_r^2\rangle = 2 [\langle\delta(x)\delta(x)\rangle -
\langle \delta(x)\delta(x+r)\rangle]$ is divergent if the 
2-point correlation function $\langle\delta(x)\delta(x+r)\rangle \simeq
r^{-\alpha}$. This problem can be handled in the same way as 
N-point correlation functions are handled, i.e., smoothing the 
density field at small scale prior to its measurement. However,
the size of the smoothing scale is put in by hand and might induce 
uncertainties, especially in studying the scaling behavior on scale 
comparable with the smoothing size. 

Ideally one would like a space-scale decomposition without the need for
pre--smoothing by hand.
This leads us to chose the discrete wavelet transform (DWT). The DWT 
performs a smoothing ``automatically" as the density difference 
$[\delta(x) -\delta(x+r)]$ on scale $r$ is calculated by a difference 
between the smoothed densities at spatial ranges $(x, x+r/2)$ and
$(x+r/2, x+r)$. In wavelet analysis, the smoothing is done 
scale-by-scale. Moreover, the bases that the DWT uses for smoothing and 
decomposing distributions are orthogonal and complete. The smoothing 
and decomposition are optimized and there is no loss of information. Furthermore, 
no mixing of the density fluctuations among different scales occurs. This 
property is excellent for studying scaling.

Here we only very briefly introduce the DWT decomposition, as it has been 
introduced in our previous publications (e.g. Pando \& Fang 1996; 
Pando \& Fang 1998; Fang \& Feng 2000). For details on the DWT refer 
to Mallat (1989a,b); Meyer (1992); Daubechies, (1992), and for physical 
applications, refer to Fang \& Thews (1998). 

We restrict our discussion to a 1-D random field of the transmission flux 
$F(x)$ extending in a spatial or redshift range $L=x_2-x_1$. To apply the DWT,
we first chop the spatial range $L=x_2-x_1$ of the 1-D sample into 
$2^j$ subintervals labeled with $l=0, ...2^j-1$ where $j$ is a positive
integer. Each subinterval spans 
a spatial range $L/2^j$. The subinterval $l$ is from $x_1+ Ll/2^j$ to
$x_1 + L(l+1)/2^j$. That is, we decompose
the space $L$ into cells $(j,l)$, where $j$ denotes the scale
$L/2^j$, and $l$ the spatial range $(x_1+Ll/2^j, x_1+L(l+1)/2^j)$.
Cell $(j,l)$ is localized in scale space and physical
(or redshift) space. 

Corresponding to each cell, there is a scaling function $\phi_{j,l}(x)$, 
and a wavelet function $\psi_{j,l}(x)$. These functions are the basis 
for the scale-space decomposition. The most important property of 
the DWT basis is its locality in both scale and physical spaces. The 
scaling function $\phi_{j,l}(x)$ is a window function on 
scale $j$ and at position $l$. The wavelet function $\psi_{jl}(x)$ is 
admissible (Daubechies 1992), i.e., $\int \psi_{jl}(x)dx=0$, and therefore, 
it measures the fluctuation on scale $j$ and at position $l$. 

With the DWT, a transmission $F(x)$ can be decomposed as (Fang \& Thews 
1998).
\begin{equation}
F(x) = \sum_{l=0}^{2^j-1}\epsilon^F_{j,l}\phi_{jl}(x)+
 \sum_{j'=j}^{J} \sum_{l=0}^{2^{j'}-1}
  \tilde{\epsilon}^F_{j',l} \psi_{j',l}(x),
\end{equation}
where $J$ is given by the finest scale (resolution) of the sample, 
i.e., $\Delta z=L/2^J$, and $j$ is the scale of interest. The scaling
function coefficient (SFC) $\epsilon^F_{jl}$ in eq.(16) is given
by projecting $F(x)$ onto $\phi_{j,l}(x)$
\begin{equation}
\epsilon^F_{j,l}=\int F(x)\phi_{j,l}(x)dx.
\end{equation}
The SFC $\epsilon^F_{j,l}$ describes the mean (or smoothed) field of 
the mode $(j,l)$.

The wavelet function coefficient (WFC), $\tilde{\epsilon}^F_{j,l}$,
in eq.(16) is obtained by projecting $F(x)$ onto $\psi_{j,l}(x)$
\begin{equation}
\tilde{\epsilon}^F_{j,l}= \int F(x) \psi_{j,l}(x)dx.
\end{equation}
The WFC is basically the difference between the smoothed flux $F(x)$ 
in cells $[x_1+ lL/2^j, x_1+ (l+1/2)L/2^j]$ and 
$[x_1+ (l+1/2)L/2^j, x_1+ (l+1)L/2^j]$. Therefore, the WFC 
$\tilde{\epsilon}^F_{j,l}$ can be 
used as the variable $\delta(x+r)-\delta(x)$ in \S 2, where  
$x \simeq x_1+ lL/2^j$ and $r \simeq L/2^j$. All compactly supported 
wavelet bases produce similar results. We will 
use Daubechies 4 in the study below. 

\subsection{The intermittent exponent in the DWT basis}

We can express eq.(3) in the DWT representation by replacing the density
difference, $\delta(x+r) - \delta(x)$, by the wavelet coefficient
$\tilde{\epsilon}^F_{j,l}$ as discussed in \S 3.1. Thus,
\begin{equation}
S^{2n}_j = \langle|\tilde{\epsilon}_{j,l}|^{2n} \rangle
=\frac{1}{2^j}\sum_{l=0}^{2^j-1}|\tilde{\epsilon}^F_{j,l}|^{2n},
\end{equation}
where $j$ plays the same role as $r$ in eq.(3). $S^{2n}_j$ 
is the mean of moment $|\tilde{\epsilon}^F_{j,l}|^{2n}$ over the
position index $l$. With the DWT, eq.(4) becomes 
\begin{equation}
S^2_j =\frac{1}{2^j}\sum_{l=0}^{2^j-1}|\tilde{\epsilon}^F_{j,l}|^2.
\end{equation}
In Paper I, $S^2_j$ is used to define the DWT power spectrum $P_j^u$,
i.e. $P^u_j=S^2_j$. $P^u_j$ is the power spectrum of the transmitted 
flux fluctuations 
$\Delta F=F -\langle F\rangle$. In Paper I, or Jamkhedkar, Bi \& Fang 
(2001), and Yang et al. (2001a), the power spectrum $P_j^u$ differs 
from eq.(20) and (21) by 
a noise term. Since the noise is gaussian, this term can be ignored on 
scales for which $S^2_j$ is larger than the variance of the noise. 

Eq.(20) can also be written as
$S^2_j =\frac{1}{2^j}\sum_{l=0}^{2^j-1}P^u_{jl}$, where
\begin{equation}
P^u_{jl}=|\tilde{\epsilon}^F_{j,l}|^2
\end{equation}
where $P^u_{jl}$ is the local power, i.e., the power on scale $j$ at 
position $l$.  

Considering $r \simeq L/2^j$, the intermittent exponent $\zeta$ [eq.(5)]
can be calculated from 
\begin{equation}
\frac{S^{2n}_j}{[S^{2}_j]^n} \propto  2^{-j\zeta}.
\end{equation}
Generally, $\zeta$ depends on $n$ and $j$.

\section{Samples}

\subsection{Keck data of Ly$\alpha$ forests}

The experimental data used in our study is the same as Paper I. It consists 
of 28 Keck HIRES QSO spectra (Kirkman \& Tytler 1997). The QSO emission 
redshifts cover a redshift range from 2.19 to 4.11. For each of the 28 
QSO's, the data are given in terms of pixels with wavelength 
$\lambda_i$, flux $F(\lambda_i)$ and noise  $\sigma(\lambda_i)$.  
The noise includes the Poisson fluctuations and the noise due to the 
background and the instrumentation. The continuum of each spectrum is 
given by IRAF CONTINUUM fitting. 

For our purposes, the useful wavelength region is from the Ly$\beta$ 
emission to the Ly$\alpha$ emission, excluding a region of about 0.06 
in redshift close to the quasar to avoid any proximity effects. In this 
wavelength range, the number of pixels is about $1.2\times 10^4$ for 
each spectrum.

For all bins in this data set, the ratio $\Delta \lambda/\lambda$ is 
constant, \ $\Delta \lambda/\lambda \simeq 13.8\times 10^{-6}$, or 
$\delta v \simeq 4.01$ km s$^{-1}$, and therefore, the resolution is 
about 8 km s$^{-1}$. The distance between $N$ pixels in units 
of the local velocity scale is given by 
$\Delta v=2c(1-\exp{[-(1/2)N\delta v/c]})$ km s$^{-1}$, or wavenumber
$k=2\pi /\Delta v$ km$^{-1}$ s. 

We do a scale by scale decomposition of the data. We use only 
$2^{13}=8192$ pixels for each spectrum. Thus, each cell on 
scale $j$ corresponds to $N=2^{13-j}$ pixels. The smallest scales are 
generally dominated by noise. We study only scales $\geq 16$ km s$^{-1}$,
corresponding to $\geq$ 4 pixels or $j \leq 11$, and 
$k \leq 0.4$ km$^{-1}$ s. Since metal lines are 
generally narrow with Doppler parameter b$<$ 15 km s$^{-1}$, ignoring scales 
less than 16 km s$^{-1}$ also suppresses metal line contamination. The 
algorithm for treating unwanted data (pixels with negative flux or missing 
data) and detected metal lines with space-scale decomposition will be 
discussed in more detail in \S 4.3.

In applying our algorithm, we sometimes use the 28 QSO transmissions 
individually, i.e.\ calculate the statistics of the transmission over 
each QSO separately, and sometimes all the transmissions are treated  
together. In the latter case, we divide the data into 12 redshift ranges 
from $z=1.6 + n\times 0.20$ to 1.6 + $(n+1)\times$ 0.20 where $n=0, 1,..11$. 
All the transmission flux in a given redshift range forms an ensemble. 
Note that the number of data points in each redshift range is different.   

\subsection{Simulation samples}

The simulated samples of the Ly$\alpha$ forests are produced 
as in Paper I. Besides the LCDM model, we also simulate the warm dark 
matter (WDM) model, for which the linear power spectrum on scales smaller 
than the free-streaming length of 
the warm particle, $R_f=0.2(\Omega_0h^2)^{1/3}(m_W/\mbox{keV})^{-4/3}$, is 
damped exponentially  with respect to the pure CDM model. 

The WDM model is believed to be a possible candidate to solve 
the problem of cuspy halos, or the singular mass profiles of massive 
objects. The problem appears in high resolution N-body 
simulations of CDM models where the models predict central cusps in the 
dark halos (Jing \& Suto 2000). However, observations instead show soft 
halo profiles as inferred from low surface brightness galaxies and the 
rotation curves of dwarf galaxies (Flores \& Primack 1994, Burkeret 1995). 
The WDM models are able to soften the density profile of the central 
core, while at the same time having  no effect on large scales. Hence, 
WDM models are proposed in order to deal with the nonlinear 
clustering on small scales. Since the cosmic field on those scales is 
already intermittent, it is important to examine the intermittent 
behavior of the WDM models. We consider four WDM models having particle 
masses $m_W=300, 600, 800$ and 1000 eV. 

\subsection{Treatment of unwanted data}

In the Keck transmission flux, there are suspect data including bad 
pixels (gaps without data), and negative flux pixels. The latter are 
generally saturated absorption regions having lower $S/N$. Although 
the percentage of low $S/N$ data is not large, it will introduce large 
uncertainties in the analysis.

As in Paper I, we use the conditional-counting method to treat 
the unwanted data. Briefly, the algorithm is as follows:
\begin{enumerate}
\item Calculate the SFCs for both the transmission $F(\lambda)$ and 
noise $\sigma(\lambda)$, i.e.
\begin{equation}
\epsilon^F_{jl}=\int F(x)\phi_{jl}(x)dx, \hspace{3mm}
\epsilon^N_{jl}=\int \sigma(x)\phi_{jl}(x)dx.
\end{equation}
\item Identify unwanted mode $(j,l)$ using the  condition 
\begin{equation}
\left| \frac{\epsilon^F_{jl}} {\epsilon^N_{jl}} \right| <  f
\end{equation}
where $f$ is a constant. This condition flags all modes with 
S/N less than $f$. We can also flag modes dominated by metal lines. 
\item Since all the statistical quantities in the DWT representation 
are based on an average over the modes $(j,l)$, we do not 
count all the flagged modes when computing these averages.  
\end{enumerate}

Condition (24) is applied at each scale $j$. If the size of a
bad data segment is $d$, condition (24) only flags modes $(j,l)$ on 
scales less than or comparable to $d$. Therefore in this algorithm,
no rejoining and smoothing of the data is needed. We also flag 
two modes around a unwanted mode to reduce any boundary effects 
of the chunks. With this method, we can still calculate 
the structure functions and intermittent exponent by eqs.(19), (20) 
and (22), but the average is not over all modes, but over the 
un-flagged modes only. Since the DWT calculation assumes the sample 
is periodized, this may cause uncertainty at the boundary. To reduce 
this effect, we drop 5 modes near the boundary. 

\section{Second and lower order statistical properties}

\subsection{The power spectrum}

The power spectra of the transmitted flux of the Keck data and simulated 
LCDM samples have been studied in detail in Paper I. The spatial 
distribution of local power $P^u_{jl}$ show spiky features. This 
property leads to large uncertainty of $P^u_j$.  

\begin{figure}
\plotone{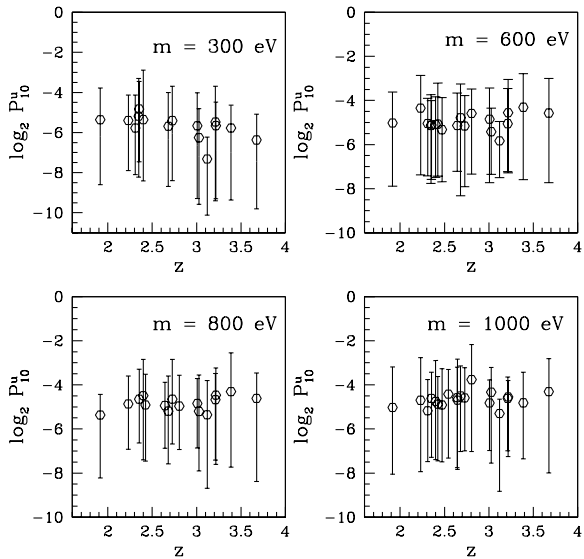}
\caption{The power, $P^u_j$, on scale $j=10$ ($k=0.2$ km$^{-1}$ s)
  of the 28 QSO's pseudo hydro simulated samples for the WDM models $m_W=300, 600,
  800$ and 1000 eV. The mean is given by average the of the power of the 20
  realizations for each QSOs. The error bars are the dispersion of 18 
  realizations, which are given by dropping the highest and lowest $P^u_j$ 
  of the 20 realizations.}
\end{figure}
Now we do a similar analysis for the WDM samples. Fig. 1 plots the 
mean and error bars for $P^u_j$ for $j=10$ 
($k = 0.2$ km$^{-1}$ s) of the 28 QSO's 
pseudo hydro simulation samples for the WDM model with $m_W=300, 600, 800$ 
and 1000 eV.  For each QSO, we calculate the mean $P^u_j$ for 
each realization then calculate the mean over the 20 realizations. 
The error bar is given by the range of $P^u_j$s by dropping 
the highest and lowest $P^u_j$ of the 20 realizations. This is 
equivalent to dropping the top and bottom 5\% of the data. 
As expected, the mean power $P^u_j$ is slightly lower for smaller $m_W$.
But the error bars of the power spectrum are generally much larger than
the $m_W$-dependence of $P^u_j$.   

\begin{figure}[h]
\plotone{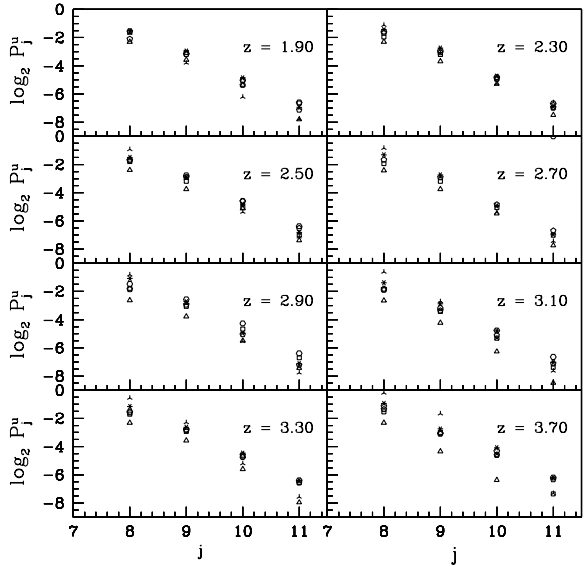}
\caption{The power spectrum $P^u_j$ vs. $j$ for Keck data 
   (vertex with three legs), LCDM (vertex with six legs) and 
   WDM models of $m_W$= 1000 (hexagon), 800 (pentagon), 600 (square) and 
   300 (triangle) eV. in redshift bins $1.90\pm 0.1$, $2.3\pm 0.1$, 
   $2.5\pm 0.1$, $2.7\pm 0.1$, $2.9\pm 0.1$, $3.1\pm 0.1$, 
   $3.3\pm 0.1$ and $3.7\pm 0.1$. The wavenumber for scale $j$ is    
   $k= 0.4 \times 2^{j-11}$ km$^{-1}$ s.}
\end{figure}
In Fig. 2, we present the mean of the power spectrum in 8 redshift bins 
and on scales $j= 8$ to 11 ($k = 0.05, 0.1, 0.2, 0.4$ km$^{-1}$ s) for 
the Keck data and all the simulated samples. On scale $j=8$ 
($k = 0.05$ km$^{-1}$ s) and for all 
redshift bins, the powers decrease in the 
order from LCDM, to the WDM of $m_W=1000$, 800, 600, and 300 eV. This order 
is expected, as 
on scale $j=8$ and redshift $z>2$, the nonlinear and intermittent features 
are weak, and the power spectra trends are about the same as in linear 
regime. That is the power is smaller for smaller $m_W$, and the LCDM 
corresponds to an infinite mass $m_W$. On scales $j>8$ 
($k > 0.05$ km$^{-1}$ s) the trend changes. This is because 
the nonlinear evolution is significant on these scales. The mean power 
is mainly dependent on rare and improbable events, and therefore, it induces 
a large uncertainty. 

Fig. 2 also shows that the power spectrum of the WDM model with 
$m_W=300$ eV is systematically lower than the power of Keck data on almost 
all scales and higher redshifts ($z>2.3$). Therefore, this model can be  ruled
out by the power spectrum. On the other hand, there are no systematic 
differences between the power spectra of the real data, and the LCDM, or 
WDM ($m_W=600, 800, 1000$eV) data. The dispersions among these power spectra 
generally are less than a factor of 2. Considering the error bars of the 
observed power spectrum 
(Paper I) and that shown in Fig. 1 are larger than by a factor of 2, the 
LCDM, $m_W=1000, 800$ and 600 eV models are basically consistent with the 
Keck data. This is one reason for proposing WDM models with $m_W > 600$ eV.

\subsection{Cumulative distribution function (CDF) of local powers}

The PDF of the local power $P^u_{jl}$ [eq.(21)] for a given $j$ are 
found to be long tailed (Paper I). This property can also be shown by the 
cumulative distribution function (CDF), which is defined as the 
percentage of the modes with local power less than a given $P^u_{jl}/P^u_j$. 
Fig. 3 presents the CDFs of $P^u_{jl}/P^u_j$ on scales $j=8, 9, 10$ 
and 11 ($k = 0.05, 0.1, 0.2, 0.4$ km$^{-1}$ s), and in redshift range 
1.7 to 1.9. 

The CDFs generally consist of two parts: a rapidly growing branch at 
$P^u_{jl}/P^u_j < 1$, and very slowly growing branch at 
$P^u_{jl}/P^u_j \gg 1$. The latter is given by spiky modes and the former 
from the passive modes between the spikes.

For a gaussian field, the PDF of $P^u_{jl}$ is a $\chi^2(N=1)$ distribution.
The CDF of the $\chi^2$ distribution is also shown in Fig. 3 (dotted 
line). This CDF approaches 1 at $P^u_{jl}/P^u_j \simeq 10$, i.e. $\sim$ 
3 $\sigma$. Fig. 3 shows that the CDF of the Keck data approaches 1 at 
$P^u_{jl}/P^u_j \simeq 15$ ($j=8$), 20 ($j=9$), 30 ($j=10$), and 60
($j=11$) and  shows that the Keck's CDF on small scales has a much longer 
tail than a gaussian field.  

\begin{figure}
\plotone{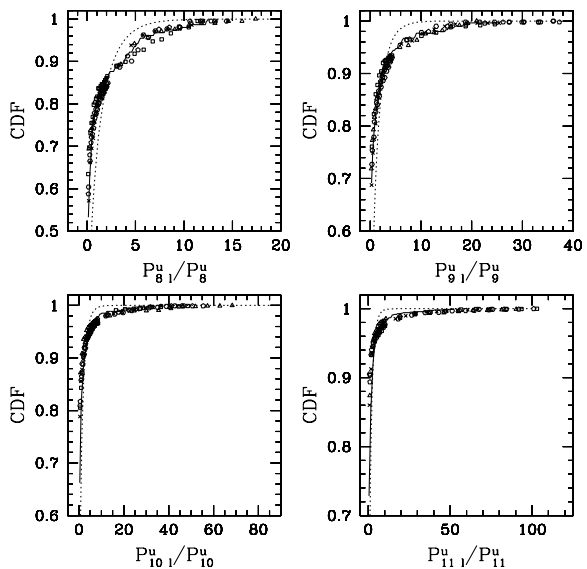}
\figcaption{The CDF of $P^u_{jl}/P^u_j$ on scales $j=8,9,10$ and 11 for 
  Keck data  
 (solid line), LCDM (cross),  and WDM models of $m_W$= 1000 (hexagon), 
  800 (pentagon), 600 (square) and 300 (triangle) eV in redshift bins 
  $1.90\pm 0.1$. The dotted lines are for gaussian field, i.e. the CDF 
  of $\chi^2(N=1)$ distribution. The wavenumber for scale $j$ is    
   $k= 0.4 \times 2^{j-11}$ km$^{-1}$ s.}
\end{figure}

We also calculate the CDFs of the quasi-hydro simulation samples for 
the LCDM and WDM models. The CDFs of simulation samples are generally 
in good agreement with the Keck data. Only the CDF of model $m_W=300$ eV 
on scale $j=8$ shows a strong deviation from Keck data at 
$P^u_{jl}/P^u_j \geq 5$.  

On scales $j=10$ and 11, the CDFs of the simulation samples show a 
little more rapid growth at $P^u_{jl}/P^u_j < 1$ than Keck data, and
longer tail than the Keck data at $P^u_{jl}/P^u_j >10$. The CDF of 
the simulation samples approaches 1 at $P^u_{jl}/P^u_j \simeq 60$ 
($j=10$), and 100 ($j=11$). This seems to indicate that the long tail 
behavior of the quasi-hydro simulation samples is more prominent than 
the observed samples. 

However, care should be taken about any conclusion made with the CDFs. 
The rapidly growing branch, which is given by the quiet 
modes, is sensitive to noise. On the other hand, the horizontal branch 
is sensitive to spiky events, and therefore, depends on the number of 
modes of the sample considered. Therefore, from the CDFs of Fig. 3,
one can only conclude that the LCDM and WDM with $m_W=600$, 800, and 
1000 eV are consistent with the Keck data.  
 
\subsection{Probability distribution functions of the SFCs}

\begin{figure}[h]
\plotone{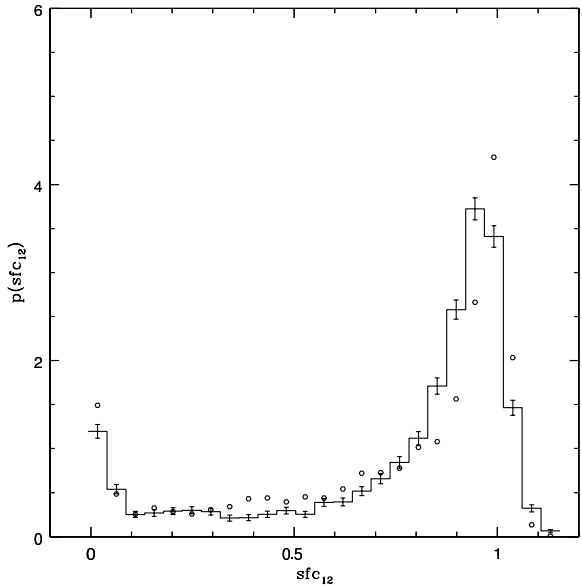}
\caption{The PDF of flux $\sqrt{2^j/L}\epsilon^F_{jl}$ on 
  scales $j=12$ for observed (histogram) and simulated LCDM samples 
  (circular) of Q0130 on scales $j=12$ ($k= 0.8$ km$^{-1}$ s). The 
  redshift of the transmitted 
  flux is in the ranges $z=2.4$ - 2.6. The error bars are given by 
  bootstrap resampling.}
\end{figure}

The PDF of the transmitted flux $F$ is often used as a statistical 
measure of the Ly$\alpha$ forests. The SFCs $\epsilon^F_{jl}$ defined 
by eq.(17) are proportional to the Ly$\alpha$ transmitted flux at 
position $l$. The $\epsilon^F_{jl}$ multiplied by $\sqrt{2^j/L}$ is 
the mean transmitted flux in the cell $(j,l)$. Therefore, the PDF of 
$\sqrt{2^j/L}\epsilon^F_{jl}$ is the PDF of $F$ smoothed on scale $j$. 

As an example, Fig. 4 presents the PDF of $\sqrt{2^j/L}\epsilon^F_{jl}$ 
for the observed QSO, Q0131 on scales $j=12$. The error bars are given by 
bootstrap resampling. In Fig. 4, we also plot the PDFs of 
$\sqrt{2^j/L}\epsilon^F_{jl}$ on $j=12$ ($k = 0.8$ km$^{-1}$ s)
for a simulation sample of the LCDM model. It can be seen, even on this 
small scale, that the PDFs of the simulation sample basically are still 
consistent with observations. The biggest difference between the observation 
and model prediction is at
$\sqrt{2^j/L}\epsilon^F_{jl}\simeq 1$. However, this difference is 
sensitive to 
noise. Fig. 5 gives the PDFs of $\sqrt{2^j/L}\epsilon^F_{jl}$ on $j=10$ 
($k = 0.2$ km$^{-1}$ s) for the LCDM simulation, but with  
gaussian noise of different $\sigma$ added. The peak at 
$\sqrt{2^j/L}\epsilon^F_{jl}\simeq 1$ is significantly dependent on the noise.
 Moreover, for the Keck sample, the noise 
level is correlated with $F$. Generally, noise is high at $F\simeq 1$,
and low at $F\simeq 0$. This makes the peak of the PDF
at $\sqrt{2^j/L}\epsilon^F_{jl}\simeq 1$ more uncertain. 

\begin{figure}[h]
\plotone{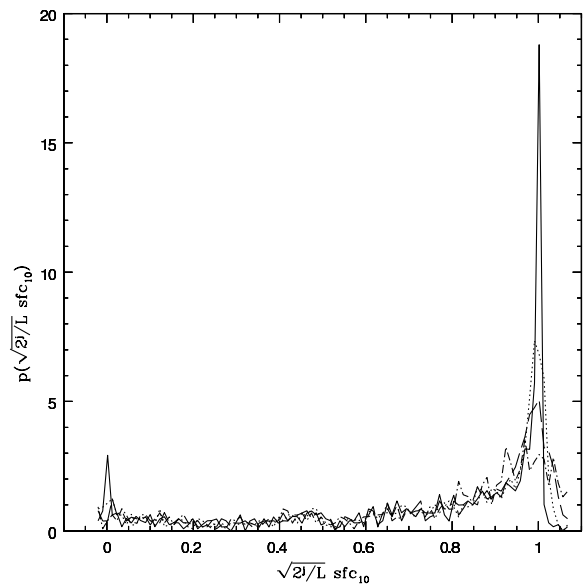}
\caption{The PDF of SFCs $\sqrt{2^j/L}\epsilon^F_{jl}$ scales 
   $j=10$ ($k= 0.2$ km$^{-1}$ s) for 
   simulated LCDM samples for Q0130. The samples are added gaussian 
   noise in each bin with standard deviation zero (solid), $\sigma$ (dot), 
   2$\sigma$ (dash), and 4$\sigma$ (dot-dash), where $\sigma$ is the 
   mean standard deviation for the keck samples.}
\end{figure}

Statistics with SFCs $\epsilon^F_{jl}$ [eq.(17)] generally are more 
sensitive to noise than WFCs $\tilde{\epsilon}^F_{j,l}$ [eq.(18)]. This is
because $\tilde{\epsilon}^F_{j,l}$ is the density {\it difference} between 
neighboring positions, and the uncertainty of the background noise on large
scales is canceled in this difference. On the other hand, $\epsilon^F_{jl}$ 
is the mean density at $l$, it is contaminated by the uncertainty of the 
background noise on all large scales. Thus, one may conclude that the 
PDFs of flux $F$ is not reliable as a discriminator.
 
\section{Structure functions and intermittent exponent 
   of Ly$\alpha$ forests}

\subsection{Structure functions}

With eqs.(19) and (22), we calculate the structure functions 
for the transmitted flux fluctuations of the Keck data. A typical result 
is plotted in Fig. 6, which is given by the Keck data in redshift range 
$z=2.50\pm 0.1$ for all QSOs. The error 
bars are the maximum and minimum of bootstrap resampling. It is well known 
that the  higher order correlation functions for large scale structure samples
have larger error than the 2-point correlation function. 
However, even for the 8th order statistic $\ln S^{8}_j/(S^{2}_j)^4$, the 
error bars are not much larger than that of the power spectrum, i.e., 
the second order statistics $P^u_j$. The structure functions are a very stable
statistical quantity because the structure functions are a 
ratio between $S^{2n}_j$ and $(S^{2}_j)^n$, which reduces the effect 
caused by high spikes. 

Fig. 6 shows that, for a given $n$, $\ln S^{2n}_j/(S^{2}_j)^n$ increases
with scale $j$, and therefore, the intermittent exponent $\zeta$ is non-zero 
and negative [eq.(22)]. The distribution of the transmitted flux fluctuations 
are neither gaussian nor self-similar, but essentially intermittent. At other 
redshift ranges, 
the structure functions behave similarly. This conclusion was already 
found with the local power distribution (Paper I). However, the local power
distribution is based on the measures of each individual mode, 
$\tilde{\epsilon}^F_{j,l}$ and has a large uncertainty. 
The distribution of local power is not effective for model discrimination. 
On the other hand,
the structure function is calculated by taking an average among modes. 
Statistically, they are more effective to test models.

\begin{figure}[h]
\plotone{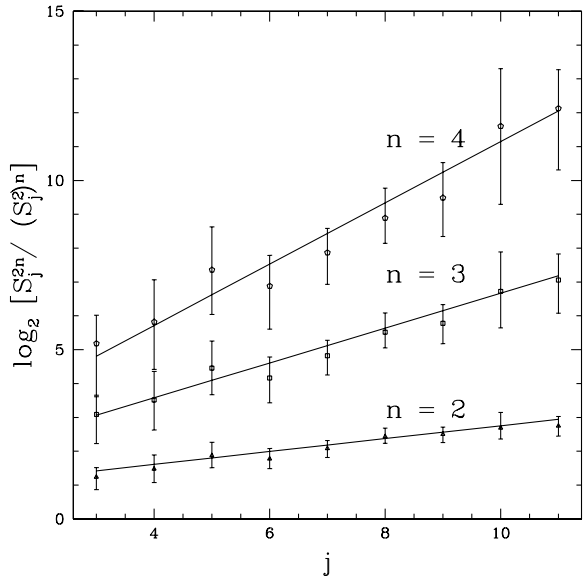}
\caption{$\log_{2} [S^{2n}_j/(S^{2}_j)^n]$ vs. j with n=2, 3, 4
     for Keck samples in the redshift range $z=2.5\pm 0.1$. The
     error bars are given by the maximum and minimum of  bootstrap 
    resampling. The wavenumber for scale $j$ is    
    $k= 0.4 \times 2^{j-11}$ km$^{-1}$ s.}
\end{figure}

This point can be seen with Fig. 7, which plots the structure function 
$\ln S^{2n}_j/(S^{2}_j)^n$ of the transmitted flux fluctuations of the 
simulation samples for the LCDM and WDM models in redshift range 
$z=2.50\pm 0.1$. The error bars are the maximum and minimum of bootstrap 
resampling. As a comparison, the structure functions of the Keck data are 
shown in each panel too. We can see from Fig. 7 that the structure 
functions of the LCDM and WDM models show intermittent behavior.
However, the $j$-dependence of the structure functions of either the 
LCDM or the WDM models is clearly different from the Keck data. For the 
simulation samples, the slope is steeper than the Keck data. This result
shows that the intermittent behavior cannot be measured by any individual 
high order moment or correlation function, but must be measured
by the scale- and $n$-dependencies of the ratio between the moments.

It is also interesting to point out that the structure functions on small 
scales $j=9$, 10, and 11 ($k=0.1, 0.2$ and 0.4 km$^{-1}$ s) are larger 
for smaller mass $m_W$. The structure 
functions of $m_W=300$ eV are always the largest one compared to the  
other models. This is because the WDM model with smaller mass $m_W$, or longer 
free-streaming length $R_f$, lacks power on small scales. The 
density fluctuations on those small scales mainly originate from the nonlinear 
process of transferring power from large to small scales (Suto \& Sasaki 
1991). On the other hand, the structure function $\ln S^{2n}_j/(S^{2}_j)^n$ 
is  the 2n-th moment $S^{2n}$ normalized by the power $S^2$, and 
hence, essentially measures the fraction of density fluctuations which 
undergoes a nonlinear evolution due to transfer of power. Therefore, 
smaller $m_W$ leads to a stronger intermittency on scales less than $R_f$.

\begin{figure}[h]
\plotone{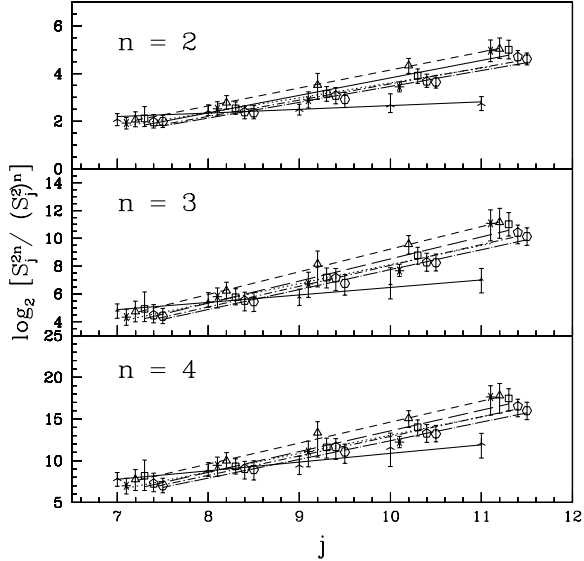}
\caption{$\log_2 [S^{2n}_j/(S^{2}_j)^n]$ vs. $j$ for the models of 
   the LCDM (star) and WDM $m_W$= 1000 (hexagon), 800 (pentagon), 
   600 (square) and 300 (triangle) eV in the redshift range $z=2.5\pm 0.1$,
   and $n=2$, 3 and 4. For comparison, the result for Keck 
   data(vertex with three legs) is also plotted. The error bars are 
   given by the maximum and minimum of bootstrap 
   resampling. For clarity, the points for simulation samples are shifted
   horizontally to the right from the corresponding $j$. The wavenumber for 
   scale $j$ is $k= 0.4 \times 2^{j-11}$ km$^{-1}$ s.}
\end{figure}

Figs.\ 6 and 7 show that the relations of $\ln S^{2n}_j/(S^{2}_j)^n$ vs. $j$
for all models and Keck data can be very well fitted by a line. That
is, the intermittent exponent $\zeta$, which is the slope of the
fitting line, is a constant in the scale range from
$j=5$ to 11 [eq.(22)], i.e. from $k= 0.006$ to 0.4 km$^{-1}$ s. 

\subsection{$n$-dependence of the intermittent exponent}

We now turn to the $n$-dependence of the structure function and 
intermittent exponent. From eq.(22), we have 
\begin{equation}
\zeta(n) = -\frac{1}{j}\ln_2 \frac{S^{2n}_j}{(S^{2}_j)^n} + {\rm const},
\end{equation}
or 
\begin{equation}
\zeta(n) - \zeta(1) = -\frac{1}{j}\ln_2 \frac{S^{2n}_j}{(S^{2}_j)^n},
\end{equation}
For a given $j$, the $n$-dependence of $\zeta(n)$ is given by 
$\ln_2 S^{2n}_j/(S^{2}_j)^n$ vs. $n$. 

Fig. 8 presents $\ln_2 S^{2n}_j/(S^{2}_j)^n$ vs. $n$ for the Keck data 
in the redshift 
range $z=2.50 \pm 0.1$. For a gaussian field, the $n$-dependence of
$\ln_2 S^{2n}_j/(S^{2}_j)^n$ is given by eq.(7), i.e. $\ln_2 (2n-1)!!$, 
which is also plotted in Fig. 8. The figure shows that the difference
between the Keck data and a gaussian field is greatest at small scales.  

More interesting is to fit the observed $n$-dependence of 
$\ln_2 S^{2n}_j/(S^{2}_j)^n$ with 
\begin{equation}
\ln_2 \frac{S^{2n}_j}{(S^{2}_j)^n} \propto n^{\alpha}(n-1).
\end{equation} 
The motivation is simple, as $\alpha=0$ corresponds to 
hierarchical clustering [eq.(12)], and $\alpha=1$ to a lognormal field 
[eq.(15)]. Fig. 8 shows that the best fit of $\alpha$ for the Keck 
data is about $0.3$ on scales $j=8$ and 9 ($k= 0.05$ and 0.1 km$^{-1}$ s), 
but $\alpha \simeq 0.5$ or higher on scales $j=10$ and 11 
($k= 0.2$ and 0.4 km$^{-1}$ s). This indicates that the transmitted 
flux is closer to a lognormal field on small scales. In other 
words, on scales for which the intermittency has been fully developed, 
the transmitted flux field can be modeled by a lognormal 
field. This may be the reason that lognormal models of Ly$\alpha$ forests 
match very well with observations not only at second and lower order 
statistics of Ly$\alpha$ forests Bi \& Davidsen, 1997), but also 
with higher order behavior, like the scale-scale correlations 
(Feng \& Fang 2000).

\begin{figure}[h]
\plotone{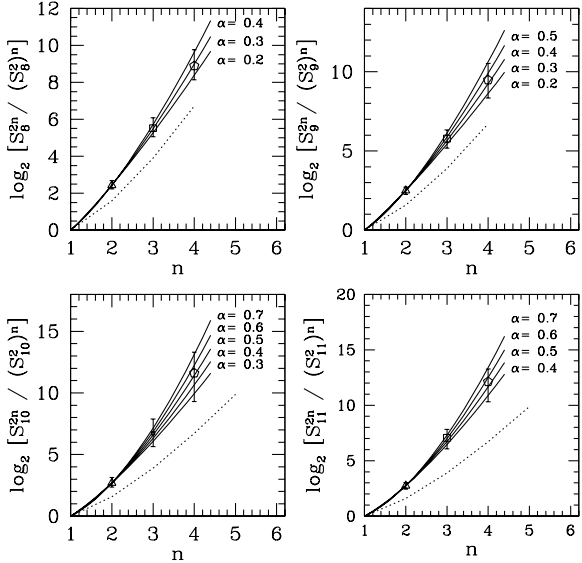}
\caption{$\log_2 [S^{2n}_j/(S^{2}_j)^n]$ vs. $n$ for $j=8,9,10$ and 
   11 for the Keck data in the redshift range $z=2.50 \pm 0.1$. The 
   error bars are given by bootstrap resampling. The fitting curves are 
   $n^{\alpha}(n-1)$. The dotted curves are for gaussian field, i.e.
   $\log_2 [S^{2n}_j/(S^{2}_j)^n]=\log_2 (2n-1)!!$.  The wavenumber for 
   scale $j$ is $k= 0.4 \times 2^{j-11}$ km$^{-1}$ s.} 
\end{figure}

Figs. 9 and 10 are, respectively, the $\ln_2 S^{2n}_j/(S^{2}_j)^n$ vs. $n$ 
for the LCDM model and 
the WDM models with $m_W=300$ eV. They are significantly different from 
a gaussian field. This is consistent with the CDFs results (\S 4.2). 
The best fit for $\alpha$ is always about 0.2, regardless of scale. For the
WDM models with other mass $m_W$, the results are similar to Figs. 9 and 10.
That is, the intermittency of the simulation samples is approximately 
that of hierarchical clustering, i.e., a mono-fractal distribution.

Thus, one may conclude that the intermittency of the simulation samples
is different from those of the Keck data not only quantitatively, but also
qualitative. The former is scale-independent, and close to a hierarchical 
clustering, while the later is close to a lognormal field on small scales. 
 
\subsection{Redshift-dependence of intermittent exponent}

\begin{figure}[h]
\plotone{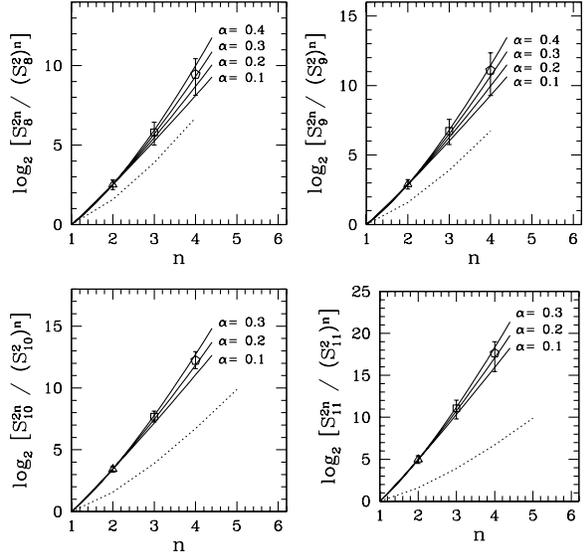}
\caption{$\log_2 [S^{2n}_j/(S^{2}_j)^n]$ vs. $n$ for $j=8,9,10$ and 11
  for the simulation samples of the LCDM in the redshift range 
  $z=2.50 \pm 0.1$. The error bars are given by the maximum and minimum of
  bootstrap resampling. 
  The fitting curves are $n^{\alpha}(n-1)$. The dotted curves are for 
   gaussian field, i.e. $\log_2 [S^{2n}_j/(S^{2}_j)^n]=\log_2 (2n-1)!!$. 
  The wavenumber for 
   scale $j$ is $k= 0.4 \times 2^{j-11}$ km$^{-1}$ s.} 
\end{figure}

\begin{figure}[h]
\plotone{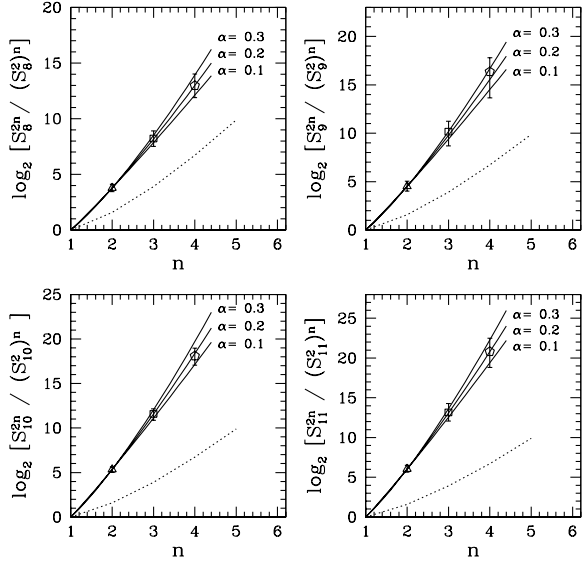}
\caption{$\log_2 [S^{2n}_j/(S^{2}_j)^n]$ vs. $n$ for $j=8,9,10$ and 11
  for the simulation samples of the WDM $m_W=300$ eV. in the redshift 
  range 
  $z=2.50 \pm 0.1$. The error bars are given by bootstrap resampling. 
  The fitting curves are $n^{\alpha}(n-1)$. The dotted curves are for 
   gaussian field, i.e. $\log_2 [S^{2n}_j/(S^{2}_j)^n]=\log_2 (2n-1)!!$.
   The wavenumber for scale $j$ is 
   $k= 0.4 \times 2^{j-11}$ km$^{-1}$ s.} 
\end{figure}

In the last two subsections, only redshift $z=2.4$ to 2.6 is considered.
For other redshift bins, the result is about the same as $z=2.4$ to 2.6. 
The $\zeta$ against redshift $z$ for $n=2$, 3 and 4 is shown in Fig. 11. 
The error bars are from the least squares fitting [eq.(25)]. 

In all redshift ranges, the value of $|\zeta|$ for the Keck data is found
to be substantially and systematically lower than that of the LCDM and WDM
models. This can directly be seen from with Fig. 7. This is also consistent 
with Figs. 8-10. It is interesting that either for the Keck data or the 
models, the intermittent exponents are almost independent of redshifts 
in the range $z= 2$ to 4. This is very different from both the observation
and theory of massive halos. Collapsed halos with mass on the order of 
galaxies and clusters undergo a significant evolution in the redshift 
range from 4 to 2.

The difference between the statistics of the Ly$\alpha$ forests and massive halos
is probably due to the fact that the QSO's transmitted flux is given mainly 
by the absorption of baryonic matter outside of the collapsed massive halos, 
i.e., the weakly clustered area. Collapsed halos correspond to the 
saturated absorption in the transmitted flux for which the S/N generally is 
low. Therefore, the Ly$\alpha$ forest does not contain information on the 
details of the massive 
halos.

Fig. 11 also shows that the intermittent exponents of the LCDM and WDM models
with $m_W=300$ eV are about the same.  This is also different from massive 
halos, for which the LCDM and WDM models predict different mass density 
profiles and number of substructures.

We have simulated the intermittent formation at high redshift. The results 
show that in the area outside of the collapsed halos, the intermittency is 
already well developed at redshift $z\sim 5$, and does not increase much for 
$z\leq 5$. Therefore, intermittency is probably the earliest developed 
nonlinear feature formed during the evolution from the linear to the 
nonlinear regimes. 

\begin{figure}[ht]
\plotone{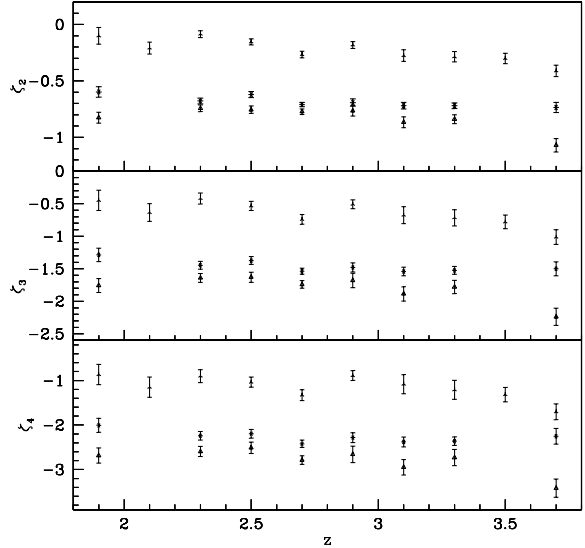}
\caption{$\zeta_n$ vs. $z$, with n=2, 3, 4 for the Keck data and 
  the LCDM (a), and WDM with $m_W=300$ eV. The error bars 
  are from the least squares fitting.}
\end{figure}

\subsection{The effect of noise}

In the above calculations, we generally take the parameter $f =1$, i.e.,
drop modes with $S/N \leq 1$. The effect of noise on the statistical 
result can be estimated by the $f$-dependence, as
large $f$ is equal to adding large noise to the simulation samples, but still
taking $f=1$. In Paper I, we showed that the power spectrum $P_j^u$ is
almost $f$-independent when $f=1$ to 5.
    
Fig. 12 shows $\ln S^{2n}_j/(S^2)^n$ vs. $j$ for the LCDM model when the 
noise and $f$ are treated as follows: 1.) sample without 
adding noise, and $f=0$; 2.) sample without adding noise and $f=3$; 3.) 
sample with adding noise and $f=3$. Fig. 12 shows that the effect of noise 
and $f$ should be considered only on the smaller scale $j=11$, and can 
be ignored on all other scales. 
 
Fig. 13 is similar to Fig. 10, but with added noise and taking $f = 3$. 
Similar to Fig. 12, the effect of noise and $f$ on $\alpha$ should be 
considered only on the smallest scale $j=11$ and gives $\alpha$ of
about 0.45. Even in this case, the simulation samples are still 
significantly different from the Keck data. Therefore, one can conclude 
that all results of \S 6.1 to 6.3 are not very sensitive to noise. 

\begin{figure}[h]
\plotone{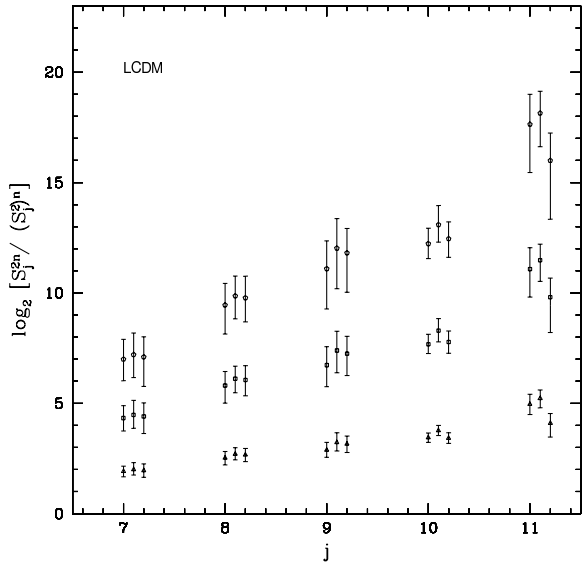}
\caption{$\log_2 S^{2n}_j/(S^2)^n$ vs. $j$ for simulation sample of 
the LCDM. The noise and $f$ are treated by the following 
ways: 1.) sample without adding noise, and $f=0$ (triangle), 2.) sample 
without adding noise and $f=3$ (square), 3.) sample with adding
noise and $f=3$ (pentagon). For clarity, the points for 2.) and 3.)
are shifted horizontally to right from the corresponding $j$. The 
wavenumber for scale $j$ is $k= 0.4 \times 2^{j-11}$ km$^{-1}$ s.  }
\end{figure}

\begin{figure}[h]
\plotone{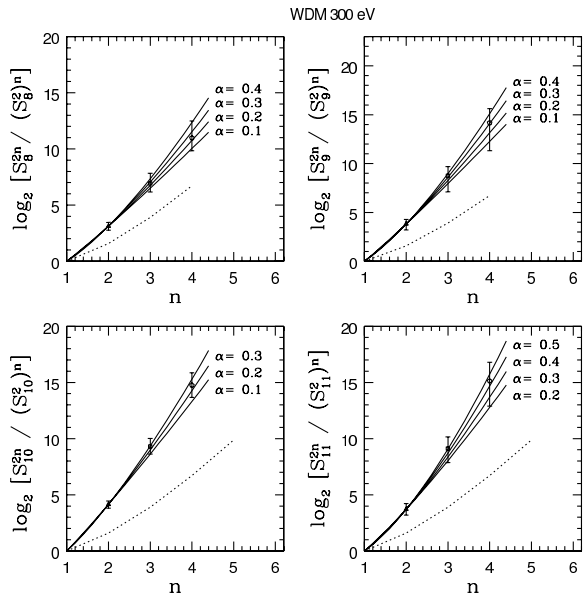}
\caption{The same as Fig.\ 10, but the sample is added with noise
and $f$ is taken to be 3.}
\end{figure}

\section{Discussions and conclusions}

\subsection{Model-discriminator of nonlinear regime}

In the Paper I, we showed that in the nonlinear regime, the cosmic mass field
is intermittent and the power spectrum  will on longer be a critical 
discriminator among models of structure formation. 
In this paper, we show that for an intermittent field, the structure 
functions and intermittent exponent are the critical discriminators 
for models of structure formation. This discriminator can be employed not 
only to distinguish gaussian and non-gaussian fields, but to detect the
{\it type} of the non-gaussianity. With the $n$- $j$- and 
$z$-dependence of the structure functions and the intermittent 
exponent, we are able to distinguish between various nonlinearly evolved fields
 in detail. That is, the intermittent exponent provides both qualitative 
and quantitative measures of the cosmic mass and velocity fields 
from the linear to nonlinear regimes and from large to small scales. 
  
We show that the distribution of the transmitted flux fluctuations 
of the QSO's Ly$\alpha$ forests is intermittent and closer to a lognormal 
field on small scales. Obviously, this conclusion is important in order to 
understand the dynamics of the clustering evolution of baryonic matter. 
It will be interesting to study when this feature formed, and  whether 
the intermittency on scales less than $j=11$ ($k=0.4$ km$^{-1}$ s) is 
closer to a lognormal field. These problems can be studied with samples 
at higher redshift and higher S/N than the one used in this work. 

Using intermittent features, even ``weakly" clustered samples, such as 
QSO's Ly$\alpha$ absorption spectrum, can play an important role for testing 
structure formation dynamics in the nonlinear regime.
It is also important to detect the intermittency of galaxy 
distributions, as the transmitted flux of Ly$\alpha$ cannot provide 
information of highly collapsed regions in the cosmic mass field. Note 
that from the definition of the structure function eq.(1), the 
intermittent exponent is bias-free if the galaxy bias is linear, i.e. 
$\rho_{galaxy}=b\rho_{dark \ matter}$. 

\subsection{Problems with the LCDM and WDM samples}

Although the pseudo-hydro simulation samples for the LCDM and WDM models 
are good in agreement with the Keck data in terms of the power spectrum, 
PDF and CDF of the transmission flux fluctuations, they are significantly and
systematically inconsistent with the intermittent features of the Keck 
data.  Specifically,  1.) the structure functions of the simulation samples 
are larger than that of Keck data on scales less than $k=0.1$ km$^{-1}$ s; 
2.) the intermittent exponents of the simulation samples are more negative 
than that of Keck data on all redshifts considered; 3.) the $n$-dependence 
of the intermittent exponent of simulation samples are close to the 
intermittency of hierarchical clustering on all scales, while the Keck 
data are close to a lognormal field on small scales. That is, the 
model-predicted intermittency is quantitatively and qualitatively 
different from the observed results. 

This is probably the first result to reveal the deviation of the popular 
dark model LCDM model from the Ly$\alpha$ forest observations. However, 
we should be very careful in reaching conclusions ruling out the relevant 
dark matter models. For a linear gaussian field, all statistics can be
determined by power spectrum with given dark matter parameters.
However, intermittency arises from the nonlinear evolution and depends 
not only on the cosmological parameters, but also on 1.) dynamical 
assumptions of the relation between the IGM and underlying dark matter
field and 2.) parameters used for simulation. 

For instance, the lognormal model of Ly$\alpha$ forests (Bi \& Davidsen
1997, Feng \& Fang 2000) assumes phenomenologically that the nonlinear field 
of baryonic matter is given by a lognormal transformation of the underlying 
mass field in linear regime. The sample given by this dynamical assumption 
can certainly fit with a lognormal PDF of the Ly$\alpha$ transmitted flux.   

In the pseudo hydro simulation, the density of baryonic matter in each pixel
is assigned according to the dark matter density at that pixel. 
This is equal to phenomenologically assuming that the statistical property of
the baryonic matter is determined by the underlying mass field. This 
assumption is reasonable in terms of second and lower order statistics.
However, it may not be correct in the context of intermittency. That is, 
although baryonic matter can be considered as a passive component in 
the system consisting of dark matter and baryonic matter, the statistical 
properties of the passive component can decouple from the underlying 
mass field during the nonlinear evolution (Shraiman \& Siggia  
2000). For instance, a diffused passive substance can exhibit intermittency, 
even when the underlying mass field is gaussian (Kraichnan, 1994). This 
is due to the nonlinear evolution of this two component system. 

Moreover, the parameters of simulation may also cause uncertainty in measuring
the intermittency. In the paper I, we studied the effect of the sample size 
on the power spectrum. When the characteristic spacing between high spikes 
(long tail events) exceed the spatial size of samples, the spatial averages 
cease to coincide with ensemble averages. As a consequence, the mean power of 
the transmission fluctuations is concentrated in rare but large spikes while 
the other modes have low power, or are even inactive. We showed that on scales 
less than $k \simeq  0.10$ km$^{-1}$ s, 50\% or more of the power is 
contributed by the top 5\% modes for unnormalized power 
spectrum, and 1\% modes for the locally-normalized power spectrum. Therefore, 
before making conclusions about the dark matter parameters, we should
study whether the deviation is caused by the dynamical assumptions on the
relation between the IGM and dark matter. We should also study the 
effects of simulation parameters (size, resolution etc) on intermittency.

Nevertheless, the current result is strong enough to conclude that 
all theory and models on the dynamics of nonlinear evolution of the 
cosmic mass and velocity fields must be examined with their predictions of 
intermittency. For models which are almost degenerate 
in the linear regime, the tests with intermittency in nonlinear 
regime are crucial. 

Finally, we should mention that the dynamical evolution of cosmic 
clustering was found to be able to sketch by the stochastic-force-driven 
Burgers' equation (Berera \& Fang 1994; Jones 1999), or the so-called KPZ 
equation (Kardar, Parisi \& Zhang 1986; Barab\'asi \& Stanley  1995). These 
equations are typical of dynamical models which lead to intermittency (e.g. 
Polyakov, 1995; E, et al 1997; Balkovsky et al. 1997). These examples 
clearly illustrate that intermittency is a basic property of the nonlinearly 
evolved cosmic mass field. Hence, we strongly believe that intermittency is 
a window to the dynamical evolution of cosmic nonlinear 
clustering. Knowledge of intermittency may ultimately lead to a better 
understanding of the still intractable problem of galaxy formation.  

\acknowledgments

LLF acknowledges support from the National Science Foundation of China 
(NSFC) and National Key Basic Research Science Foundation. 
PJ also acknowledges the support of the Dean's Dissertation Fellowship 
at the University of Arizona. We also thank Stephane Colombi for some very
insightful comments that improved this manuscript.

\end{document}